\documentclass[twocolumn,shortnote]{jpsj2} 
\title{Tunnelling spectroscopy of the interface between 
Sr$_2$RuO$_4$ and a single Ru micro-inclusion in eutectic crystals}

\author{
	Hiroshi \textsc{Yaguchi}$^{1}$, 
	Keiichi \textsc{Takizawa}$^{1}$, 
	Minoru \textsc{Kawamura}$^{2,3}$, 
	Naoki \textsc{Kikugawa}$^{1}$,
	Yoshiteru \textsc{Maeno}$^{1,4}$, 
	Takayoshi \textsc{Meno}$^{5}$,
	Tatsushi \textsc{Akazaki}$^{2}$,
	Kouichi \textsc{Semba}$^{2}$
	and Hideaki \textsc{Takayanagi}$^{2}$
}

\inst{
	$^{1}$Department of Physics, Graduate School of Science,
		Kyoto University, Kyoto 606-8502\\
	$^{2}$NTT Basic Research Laboratories,
		NTT Corporation, Atsugi, Kanagawa 243-0198 \\
	$^{3}$RIKEN, Wako, Saitama 351-0198\\
	$^{4}$Kyoto University International Innovation Center,
		Kyoto 606-8501\\
	$^{5}$NTT-ATN Corporation, Atsugi, 
	Kanagawa 243-0018
}

\kword{Sr$_2$RuO$_4$, tunnelling spectroscopy, zero bias conductance peak, spin-triplet superconductivity, ruthenate}

\begin{document}
\maketitle

The understanding of the zero bias conductance peak (ZBCP) 
in the tunnelling spectra of S/N junctions involving  {\it d}-wave cuprate superconductors 
has been important in the determination of the phase structure 
of the superconducting order parameter~\cite{ZBCPtheory}. 
In this context, the involvement of a $p$-wave superconductor 
such as Sr$_2$RuO$_4$ ~\cite{discovery} 
in tunnelling studies is indeed of great importance. 
We have recently succeeded in fabricating devices 
that enable S/N junctions forming at interfaces 
between Sr$_2$RuO$_4$ and Ru micro-inclusions in eutectic crystals
to be investigated~\cite{kawamura}. 
We have observed a ZBCP and have interpreted it 
as due to the Andreev bound state, commonly seen in unconventional superconductors. 
Also we have proposed that the onset of the ZBCP may be used 
to delineate the phase boundary for the onset of a time reversal symmetry broken (TRSB) state 
within the superconducting state, 
which does not always coincide with the onset of the superconducting state~\cite{kawamura}. 
However, these measurements always involved two interfaces 
between Sr$_2$RuO$_4$ and Ru. In the present study, 
we have extended the previous measurements to obtain 
a deeper insight into the properties of a single interface 
between Sr$_2$RuO$_4$ and Ru.

The layered peroviskite Sr$_2$RuO$_4$, 
recognised as one of the strongest candidates for a spin-triplet superconductor, 
has attracted great research interest in spite of its relatively low $T_{\rm c}$ of 1.5 K \cite{review}. 
Taken together several experiments, such as NMR \cite{NMR} and muon spin relaxation \cite{muSR}, 
the basic form of the vector order parameter is constrained to be 
{\bf{\textit{d}}}(\textbf{\textit{k}}) = {\bf{\textit{z}}}$\Delta_{0}(k_x + ik_y)$, 
corresponding to a TRSB state (the chiral state). 
Although this basic form is too simplified to explain other existing experimental results, 
the incorporation of band-dependent gap structure 
and strong in-plane anisotropy of the superconducting gap allows 
the existing experimental results to be reconciled \cite{Cp}.

Interesting aspects of Sr$_2$RuO$_4$ include superconductivity 
in eutectic systems such as Sr$_2$RuO$_4$-Ru \cite{3K} 
and Sr$_2$RuO$_4$-Sr$_3$Ru$_2$O$_7$ \cite{Fittipaldi}. 
In Sr$_2$RuO$_4$-Ru, superconductivity with an enhanced $T_{\rm c}$, 
called the 3-K phase, is known to occur within Sr$_2$RuO$_4$ 
at interfaces between Sr$_2$RuO$_4$ and Ru micro-inclusions. 
Besides this, these eutectic systems, as a consequence of the eutectic solidification, 
contain natural interfaces with the spin-triplet superconductor Sr$_2$RuO$_4$. 
Under certain circumstances, such interfaces may be regarded 
as superconductor/normal metal (S/N) junctions. 
In fact, Mao  {\it et al.} performed break-junction experiments on Sr$_2$RuO$_4$-Ru eutectic 
and observed a ZBCP characteristic of unconventional superconductivity \cite{3KZBCP}.

Recently, as mentioned earlier, we have also used interfaces 
in eutectic Sr$_2$RuO$_4$-Ru to obtain well-defined S/N junctions illustrated in Fig.~1, 
using a micro-fabrication technique. 
Details of the fabrication and the crystal growth are described in Refs. 3 and 8. 
As depicted in Fig.~1, each S/N junction possesses a Ti/Au electrode 
directly attached to the Ru micro-inclusion through 2~${\mu}$m ${\times}$ 3~${\mu}$m rectangular contact windows. 
The SiO$_2$ film is an insulating layer deposited 
on the $ab$-plane of Sr$_2$RuO$_4$. 
Achieving good electrical contacts is possible practically only on Ru inclusions 
because of a non-superconducting layer 
with a high resistivity forming on the $ab$-plane surface of Sr$_2$RuO$_4$. 
Consequently, actual measurement current is injected via two electrodes on Ru inclusions. 
Therefore, the two Sr$_2$RuO$_4$/Ru junctions in series 
are inevitably involved for the measurement. 
Besides, this measurement configuration leads to the resistances of the S/N junctions, 
Sr$_2$RuO$_4$, Ru and Ti/Au electrodes in series 
all contributing to the whole resistance measured. 
However, the resistance measured is dominated 
by that of the Sr$_2$RuO$_4$/Ru junctions 
because the resistance measured, typically on the order of 1~${\Omega}$, is 
much larger than the resistance of the other components. 
We use the identical devices also in the present study. 
A lock-in technique was employed to measure the differential conductance.

\begin{figure}[tb]
\begin{center}
\includegraphics[width=35mm]{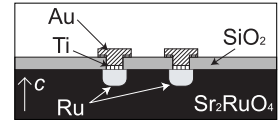}
\end{center}
\caption{Cross-sectional schematic of S/N junctions of Sr$_2$RuO$_4$ 
and Ru with Ti/Au electrodes attached on the Ru inclusions. 
Measurement current flows between the two electrodes through the two S/N junctions.}
\label{f1}
\end{figure}

Similarly to work in Ref. 10, we observed a clear ZBCP in the previous study in Ref. 3. 
The ZBCP is attributed to a sign change in the order parameter on the Fermi surface, 
characteristic of unconventional superconductivity. 
Figure 2 shows two spectra of different devices at 0.6 K. 
The upper trace was obtained using the same device as used in the previous study~\cite{kawamura}, 
and shows a sharp ZBCP in addition to a broad peak. 
By contrast, the lower trace appears to be similar but without a sharp ZBCP. 
Such S/N junctions may be phenomenologically modelled using a steep barrier at the interface, 
assuming a ${\delta}$-function potential with its strength $Z$. With increasing $Z$, 
the tunnelling limit, where the resultant spectra 
represent the density of states at the interface, is approached \cite{BTK}. 
In this context, the overall similarity of the two spectra 
and the moderate closeness of the normal conductance values 
may suggest the proximity of interface conditions 
such as the barrier height (probably rather high $Z$), 
whereas the origin of their difference is unclear. 
Based on this assumption, these observations perhaps signal that the spectrum 
depends on the (effective) direction of the interface relative to the crystallographical axes. 
In fact, S/N junctions including high $T_{\rm c}$ cuprates with {\it d}-wave pairing show 
strong dependence of the tunnelling spectrum on the direction of the interface. 
It is well established that the height of ZBCP takes a maximum (minimum) 
for $\alpha$ = $\pi$/8 ($\alpha$  = 0), where  $\alpha$  is the angle between the normal to the interface 
and the direction of a lobe of the {\it d}-wave order parameter \cite{ZBCPtheory,ZBCPexpt}. 
This was first theoretically suggested \cite{ZBCPtheory} and later demonstrated 
by well-controlled experiments \cite{ZBCPexpt}. 
It should be noted that the phase associated with the chiral state 
changes continuously between 0 and 2$\pi$, 
while the phase of the superconducting order parameter 
of high $T_{\rm c}$ cuprates takes either 0 or $\pi$.

Although the basic form of the superconducting order parameter for Sr$_2$RuO$_4$ is 
{\bf{\textit{d}}} = {\bf{\textit{z}}}$(k_x + ik_y)$ and isotropic, the gap function is considerably anisotropic in reality. 
A realistic gap function (on the $\gamma$ band) is considered to be {\bf{\textit{d}}} = {\bf{\textit{z}}}$[\sin(k_x) + i\sin(k_y)]$ \cite{Cp}, 
possesing strong in-plane anisotropy. In fact, the isotropic gap 
associated with {\bf{\textit{d}}} = {\bf{\textit{z}}}$(k_x + ik_y)$ leads to a broad ZBCP \cite{ZBCPSRO}, 
so that the observed sharp ZBCP has not been well explained. 
Further theoretical studies are awaited to understand the detailed spectra 
with more realistic band and superconducting gap structures incorporated.

\begin{figure}[tb]
\begin{center}
\includegraphics[width=75mm]{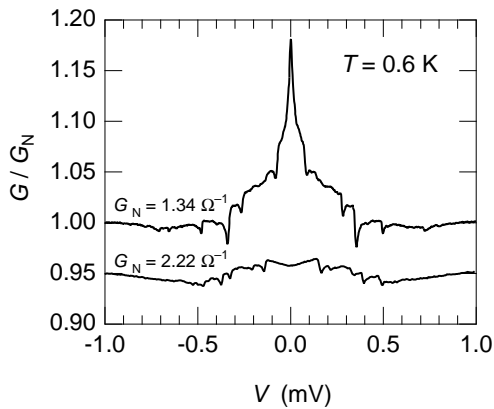}
\end{center}
\caption{Normalised differential conductances of two devices at $T$ = 0.6 K 
and $H$ = 0 as a function of the bias voltage. The upper trace shows 
a zero bias conductance peak whilst the other does not. 
Traces have been offset for clarity.}
\label{f1}
\end{figure}

Finally, we discuss the width of the superconducting gap observed. 
The study in Ref. 10 and our previous and present studies employ eutectic Sr$_2$RuO$_4$-Ru 
and thus involve 3-K phase superconductivity \cite{3K}, 
which occurs on the Sr$_2$RuO$_4$ side of the interface. 
While the spectra obtained in our present and previous studies are 
very similar to those previously reported on experiments 
using a break junction technique in Ref. 10, 
the width of the whole conductance peak in the present study 
is considerably smaller than that reported \cite{3KZBCP}. 
We suggest that this quantitative difference between those works 
may be attributed to the difference in the number of junctions involved effectively. 
While our measurements inevitably involve two junctions 
in series from the geometrical configuration, 
the work in Ref. 10 may involve several interfaces in parallel and/or series.

We have confirmed that only one of the two junctions, 
{\it i.e.} a single interface of Sr$_2$RuO$_4$ and Ru, 
is responsible for the ZBCP shown in Fig. 2 in the following way. 
We made three sets of measurements with a third junction involved 
and obtained the (differential) resistance between two junctions $R_{ij}$. 
We here label the resistance between the $i$-th and $j$-th junctions $R_{ij}$ 
and the resistance of the $i$-th junction $R_{(i)}$. 
(The upper trace in Fig. 2 corresponds to $R_{12}$.) 
As mentioned earlier, the resistance measured is dominated 
by the resistance of the Sr$_2$RuO$_4$/Ru junctions. 
Consequently, the use of the assumption $R_{ij} =  R_{(i)} + R_{(j)}$ {\it etc}. is reasonable, 
and allows the resistances between junctions to be decomposed 
into the resistance of each junction $R_{(i)}$. 
{\it e.g.} $R_{(1)} = (R_{12} + R_{23} + R_{31})/2 - R_{23}$. Figures 2 and 3 show 
the resultant resistance of each junction. 
Only one of the three junctions is responsible for the observed ZBCP~\cite{Jc}. 
Based on $R_{(1)}$ in Fig. 3, the simple attribution of the width of the broad peak to the gap width 
yields 2$\Delta$ = 0.7 meV (at 0.6 K), corresponding to 
2${\Delta}/k_{\rm B}T_{\rm c}~ {\sim}~3 ~(T_{\rm c} ~{\sim}$ 2.7 K ~\cite{kawamura}). 
(The spectrum is fairly flat in the range $|V| > 0.35$ ~meV, in support of the above attribution ie 2$\Delta$ = 0.7 meV.)
The broad peak persists well above $T_{\rm c}$ of Sr$_2$RuO$_4$, indicative of the gap 
being associated with the 3-K phase \cite{kawamura,3KZBCP,LT24} .

\begin{figure}[tb]
\begin{center}
\includegraphics[width=75mm]{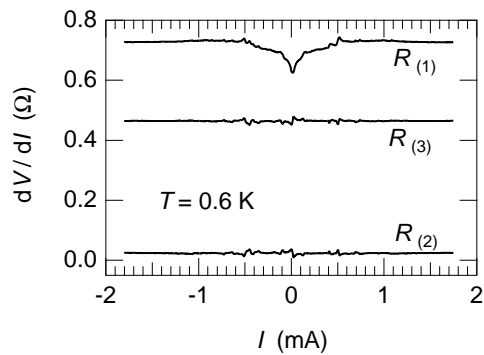}
\end{center}
\caption{Decomposed spectra as a function of current. 
Only one of the three interfaces is responsible for the clear ZBCP observed 
while the other junctions hardly show such a peaked feature. (For details, see text.) }
\label{f1}
\end{figure}

\section*{Acknowledgment} This work was supported in part by Grants-in-Aid for Scientific Research from the JSPS and 21COE program on ÒCenter for Diversity and Universality in PhysicsÓ from the MEXT of Japan.

\end{document}